\documentclass{aa}

\usepackage{graphicx}
\usepackage{txfonts}

\begin{document}

   \title{ On measuring the Tully-Fisher relation at $z > 1$\thanks{Observations carried out using the Very Large Telescope at the ESO Paranal Observatory, Program ID 68.A-0243 and 70.A-0304.}}
   \subtitle{A case study using strong H$\alpha$ emitting galaxies at $z\sim 1.5$}

   \author{L. van Starkenburg
          \inst{1}
	  \and
          P. P. van der Werf \inst{1}
	  \and
	  L. Yan\inst{2}
	  \and
          A. F. M. Moorwood\inst{3}
          }

   \offprints{}

   \institute{Sterrewacht Leiden, P.O. Box 9513, 2300 RA  Leiden, The Netherlands \\
              \email{vstarken, pvdwerf@strw.leidenuniv.nl }
         \and
              Spitzer Science Center, California Institute of Technology, MS 220-6, Pasadena, CA 91125, USA\\
             \email{lyan@ipac.caltech.edu}
          \and
	      ESO, Karl-Schwarzschild-Strasse 2, 85748 Garching bei M\"unchen, Germany\\
	      \email{amoor@eso.org}
             }

   \date{Received <date>; accepted <date>}

   \abstract{The evolution of the line width - luminosity relation for spiral galaxies, the Tully-Fisher relation, strongly constrains galaxy formation and evolution models. At this moment, the kinematics of $z>1$ spiral galaxies can only be measured using rest frame optical emission lines associated with star formation, such as H$\alpha$ and [\ion{O}{iii}]5007/4959 and [\ion{O}{ii}]3727. This method has intrinsic difficulties and uncertainties. Moreover, observations of these lines are challenging for present day telescopes and techniques. Here, we present an overview of the intrinsic and observational challenges and some ways way to circumvent them. We illustrate our results with the HST/NICMOS grism sample data of $z \sim 1.5$ starburst galaxies.  The number of galaxies we can use in the final Tully-Fisher analysis is only three.  We find a $\sim 2$ mag offset from the local rest frame B and R band Tully-Fisher relation for this sample. This offset is partially explained by sample selection effects and sample specifics. Uncertainties in inclination and extinction and the effects of star formation on the luminosity can be accounted for. The largest remaining uncertainty is the line width / rotation curve velocity measurement. We show that high resolution, excellent seeing integral field spectroscopy will improve the situation. However, we note that no flat rotation curves have been observed for galaxies with $z>1$.  This could be due to the described instrumental and observational limitations, but it might also mean that galaxies at $z>!$ have not reached the organised motions of the present day.

   } 

   \maketitle

%

\section{Introduction}

The Tully-Fisher relation (hereafter TFR) is a tight empirical relation between the flat rotation curve (RC) velocity and the luminosity of spiral galaxies  (Tully \& Fisher \cite{tullyfisher}). The TFR has been used as a distance estimator and for measurements of the Hubble constant H$_0$ (e.g. Tully \& Pierce \cite{tullypierce}). 

In addition to its empirical applications, the TFR is interesting in itself because it defines a tight relation between the total (dark matter dominated) mass of spiral galaxies and their luminosity produced by baryons. Furthermore, assuming a stellar mass-to-light ratio $M_*/L$, the stellar masses of galaxies can be calculated from their luminosities and a stellar mass TFR can be derived (Bell \& De Jong \cite{bell}). After addition of the gas mass, one obtains the baryonic mass TFR (Verheijen \cite{verheijen}; Bell \& De Jong \cite{bell}; McGaugh et al. \cite{mcgaugh}). The variations in the dark-to-baryonic mass ratio of galaxies are small and deviations from the baryonic TFR are absent down to very low mass galaxies (McGaugh et al. \cite{mcgaugh}) although others claim a slight deviation for dwarf spirals (Stil \& Israel \cite{stil}). Semi-analytical models of galaxy formation struggle to explain simultaneously the slope, zero point and tightness of the TFR in all optical and near infrared bands (Van den Bosch \cite{bosch}). 

The tight fundamental relation between mass and luminosity is interesting to study in the context of galaxy evolution. The study of the evolving TFR with redshift can provide valuable information on the luminosity evolution of galaxies and the buildup of stellar mass as a function of galaxy mass. As we will show in this paper, the analysis of high redshift TFRs needs careful treatment of observational limits, selection effects, sample definitions and starburst influences, and high resolution high quality spectra.

In the local universe, HI is used to measure the velocity profiles of spiral galaxies. The gas disk in spiral galaxies extends 2-3 times further out than the stellar disk. HI measurements are currently limited to low redshift, beyond redshift $\sim 0.2$ HI emission has not been observed (Zwaan et al. \cite{zwaan}) and one has to rely on other kinematic tracers. Another gas tracer could be CO, but CO detections in the high redshift universe are currently limited to very CO bright galaxies, like submillimeter galaxies, quasi stellar objects (QSO) and radio galaxies (Hainline et al. \cite{hainline} and references therein). One gravitationally lensed Lyman Break Galaxy (LBG) has been detected in CO (Baker et al. \cite{baker04a}).  A second attempt to detect a LBG in CO failed, although the attempt was on the dustiest LBG known (Baker et al. \cite{baker04b}). 

Bright optical narrow emission lines like H$\alpha$, H$\beta$, [\ion{O}{iii}]5007/4959 and [\ion{O}{ii}]3727 can also be used to trace the rotation curve of galaxies. Their presence is limited to the stellar disk (or more precisely: the star forming disk). In the local universe, the agreement between HI and HII measurements of RCs is excellent (Courteau \cite{courteau}). However, these lines shift out of the optical regime at redshifts 0.4 - 1.4. In recent years, high resolution near infrared spectrographs like the Infrared Spectrometer And Array Camera (ISAAC) at the Very Large Telescope (VLT) of the European Southern Observatory (ESO) have become available, opening the window out to redshift 2.4 (for H$\alpha$) TFR studies.  Examples are Rigopoulou et al. (\cite{rigopoulou}), who studied massive $z\sim0.6$ galaxies, Barden et al. (\cite{barden}) who found an offset from the local B band TFR of $\sim 1 \textrm{mag}$ at $z\sim 0.9$, Lemoine-Busserolle et al. (\cite{lemoine}) who used gravitational lenses to study two galaxies at $z\sim 1.9$ and Pettini et al. (\cite{pettini}), who studied Lyman Break Galaxies at $z\sim3$.   We used ISAAC to study the kinematics of a sample of $z\sim1.5$ H$\alpha$ emitting galaxies and we present the results in this paper as a case study for $z > 1$ TFR studies.

Our sample is a subsample of the McCarthy et al. (\cite{mccarthy}) HST/NICMOS grism survey sample. McCarthy et al. (\cite{mccarthy}) surveyed 64 square arc minutes with the slitless NICMOS G141 grism and detected 33 emission line objects with varying 3$\sigma$ limiting line fluxes down to $1\times10^{-17} \textrm{erg s}^{-1} \textrm{cm}^{-2}$. They argue that the detected emission lines are H$\alpha$ between redshift 0.75 and 1.9. The H$\alpha$+[\ion{N}{ii}]6548/6584 complex is not resolved due to the low spectral resolution ($R\sim150$) of the grism and therefore contamination by other emission lines (particularly [\ion{O}{iii}]5007/4959) cannot be excluded and no kinematic information is obtained. This sample is biased to galaxies with large H$\alpha$ equivalent width, EW(H$\alpha$), and H$\alpha$ flux, F(H$\alpha$), due to the low spectral resolution  of the grism. We chose this sample because it has clear selection criteria and all sources have known H$\alpha$ fluxes. 

Here, we present observations of 9 objects from the McCarthy et al. sample with the ISAAC at the VLT in medium resolution mode ($R\sim3000-5000$). Our aim was to resolve the H$\alpha$+[\ion{N}{ii}] complex (or the [\ion{O}{iii}]5007/4959 doublet) to confirm redshifts, measure accurate line fluxes and linewidths and if possible also rotation curves. We use this data to present our case study for $z>1$ TFRs.

Hicks et al. (\cite{hicks}) also performed a follow-up of the HST/NICMOS grism sample. They observed 14 objects aiming to detect emission lines, particularly [\ion{O}{ii}]3727 in the optical (R/I band) using LRIS at the Keck telescope. They observed in low resolution mode ($R\sim350-700$ depending on the grating used) and therefore did not obtain any kinematic information. They confirmed the redshift from McCarthy et al. (1999) for 9 out of 14 objects. They explained the non-confirmations by twilight observations or the presence of a nearby bright star (emission lines may very well be not bright enough to detect in these two cases). In two cases, the [\ion{O}{ii}]3727 line was outside the observed wavelength range and other emission lines like for example \ion{C}{ii}]2326, \ion{C}{iii}]1909 and \ion{Mg}{ii}2800 might have been too faint to detect. The fifth non-detection was explained by reddening or the emission line detected by McCarthy et al. was not H$\alpha$ but H$\beta$ or [\ion{O}{iii}]5007/4959. In the latter case, no bright emission lines are expected in the wavelength range observed.

Our follow-up is complementary in two ways: we observe the objects accessible from the southern hemisphere whereas Hicks et al. observed from the northern hemisphere. Only one object (J0931-0449) is in both samples. Second, we observe at higher resolution, resolving the emission lines. 

This paper is organised as follows. The first part of the paper describes the case study data set: the observations of the NICMOS grism sample (Section 2), data reduction and analysis (Section 3), the sample properties (Section 4) and notes on individual galaxies (Section 5). The second part of the paper discusses high redshift TFRs using the earlier discussed dataset as an illustration with a strong focus on pitfalls (Section 6). We conclude with a summary and conclusion in the final section (Section 7). 

Throughout this paper, we assume $\Omega_M=0.3$, $\Omega_\Lambda=0.7$ and $H_0=70 \textrm{ km s}^{-1}\textrm{Mpc}^{-1}$. All magnitudes in this paper are Vega magnitudes.


\section{Sample selection and observations}

The McCarthy et al. (\cite{mccarthy}) catalog contains 33 galaxies with redshifts between 0.75 and 1.9. We selected all targets from the McCarthy et al. (\cite{mccarthy}) sample accessible with the VLT and with the line falling in the J or H atmospheric window. We did not select on morphology or emission line flux. 

The observations were done in two runs. In the nights of February 23, 24 \& 25 2002 (ESO program ID 68.A-0243(A)) we observed 7 targets in visitor mode using the VLT ISAAC long slit spectrograph in medium resolution (MR) mode with $1\arcsec$ slit under varying atmospheric conditions. In winter 2003 (ESO program ID 70.A-0304(A)) two targets were observed in service mode under excellent seeing and sky conditions (seeing $<0\farcs6$, clear/photometric) with the $0\farcs6$ slit. Integration times varied between 120 to 250 minutes, depending on the atmospheric conditions and the emission line flux. One target could not be acquired, although it was attempted several times. An overview of all observations is given in Table \ref{Observations}.

\begin{figure*}
   \centering
\includegraphics[width=8.5cm]{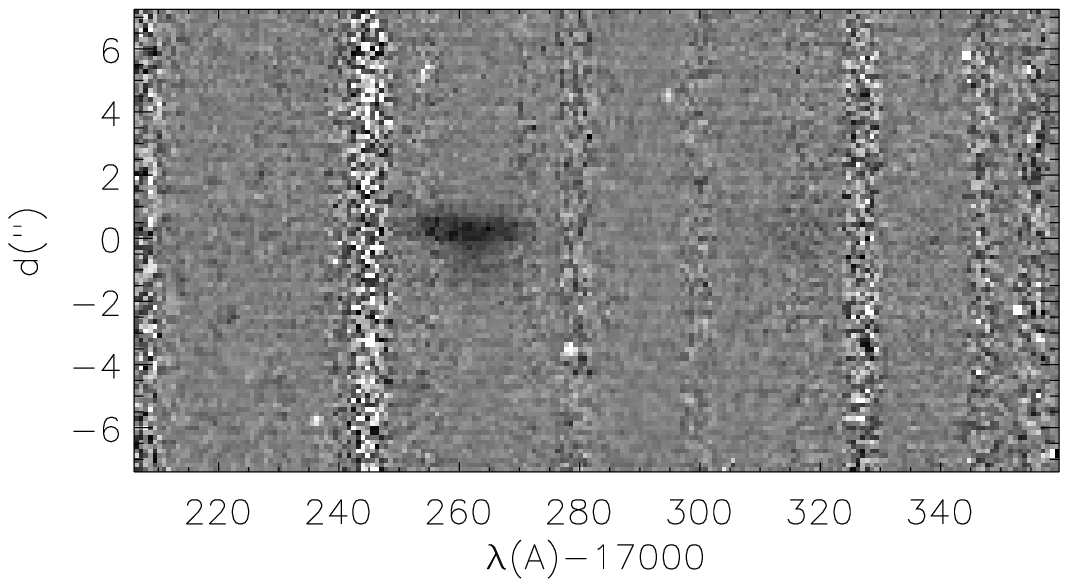}
\includegraphics[width=8.5cm]{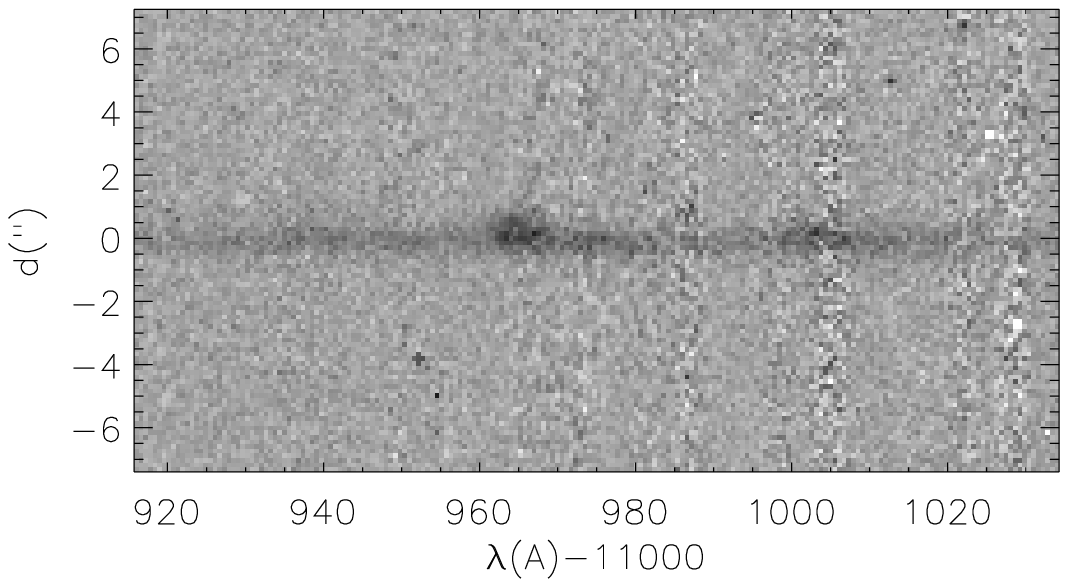}
\includegraphics[width=8.5cm]{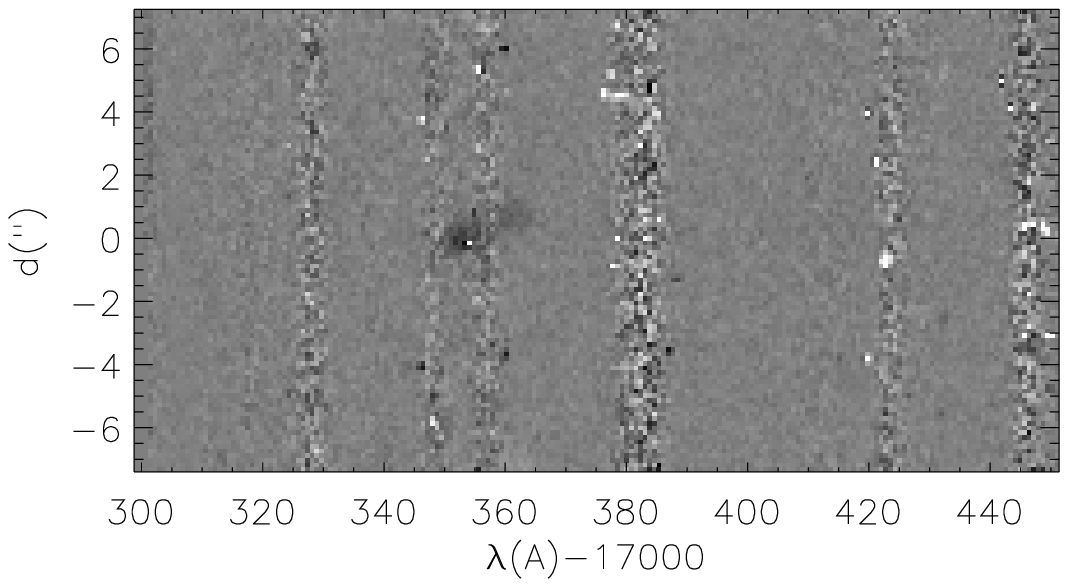}
\includegraphics[width=8.5cm]{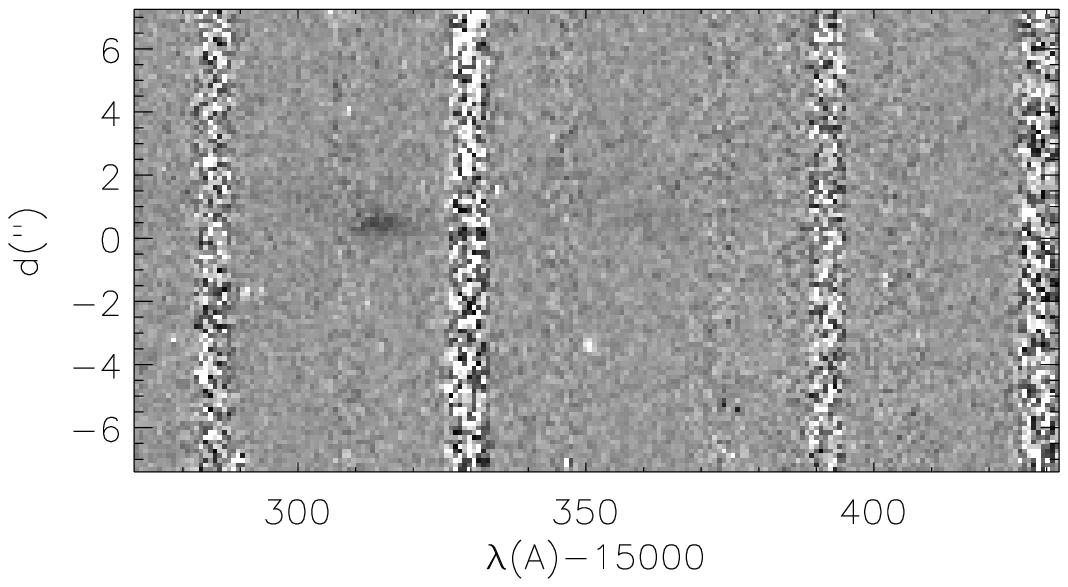}

      \caption{Two dimensional spectra, from the upper left and then clockwise: J0627-6512, J0738+0507a, J1143--8036a and J0738+0507b. Note the extended emission in J0627-6512 and J0738+0507b and the positions of the OH sky lines in all spectra. 
              }
         \label{spec2D}
   \end{figure*}

   \begin{table}
      \caption[]{Observations}         \label{Observations}
     $$ 
         \begin{array}{llccc}
            \hline
            \noalign{\smallskip}
            \mbox{Source ID}      &  \mbox{run ID} & \mbox{slit width} &   T_\mathrm{int}(\mbox{s}) &  seeing^{\mathrm{b}}    \\
            \noalign{\smallskip}
            \hline
            \noalign{\smallskip}
            \mbox{J0627--6512}  &   \mbox{68.A--0243(A)}  &  1\arcsec  &  7200   & 0\farcs69 \\
	    \mbox{J0738+0507a}  &   \mbox{68.A--0243(A)}  &  1\arcsec  &  7200   & 0\farcs83\\
	    \mbox{J0738+0507b}  &   \mbox{70.A--0304(A)}  &  0\farcs6  &  15000  & 0\farcs60\\
	    \mbox{J0931--0449}  &   \mbox{68.A--0243(A)}  &  1\arcsec  &  7200   & 0\farcs80\\
	    \mbox{J1056--0337}  &   \mbox{68.A--0243(A)}  &  1\arcsec  &  14400  & 0\farcs67\\
	    \mbox{J1143--8036a}^{\mathrm{a}} &   \mbox{68.A--0243(A)}  &  1\arcsec  &   7200 & 0\farcs78 \\
	    \mbox{J1143--8036b}^{\mathrm{a}} &   \mbox{68.A--0243(A)}  &  1\arcsec  &   7200 & 0\farcs78 \\
	    \mbox{J1143--8036c} &   \mbox{68.A--0243(A)}  &  1\arcsec  &  7200   &  0\farcs91 \\
	    \mbox{J1143--8036d} &   \mbox{70.A--0304(A)}  &  0\farcs6  &  12000  & 0\farcs60 \\
            \noalign{\smallskip}
            \hline
         \end{array}
     $$ 
\begin{list}{}{}
\item[$^{\mathrm{a}}$] J1143-8036a and J1143-8036b were observed in the same slit. 
\item[$^{\mathrm{b}}$] The seeing was measured on the brightest object in the slit in the reduced image. In one case, J1056-0337,  there was no bright object in the slit and the seeing was measured on the standard star for flux calibration. 
\end{list}
   \end{table}

The observational set-up was as follows. The slit was aligned along the major axis of the galaxy as determined from NICMOS H band images. Where possible without deviating more than 10\degr ~from the major axis of the galaxy, a bright reference star was also included in the slit to make sure the slit was on target. To facilitate sky subtraction, total integration times were dived in 12 or 15 minutes exposure times, nodding in ABBA cycles along the slit. After observation of each object, a bright nearby standard B star was observed with the same instrument setup and the same air mass to allow accurate flux calibration. For the 0\farcs6 slit observations, the B star was also observed with the 2\arcsec slit to calculate the (wavelength dependent) slit loss correction. Depending on wavelength and slit width, the sampling was 0.57 to 0.81 \AA ~pix$^{-1}$. The full width half maxima (FWHM) of the sky lines varied between 3.3 and 4.7 \AA.

\section{Data reduction and analysis}
We used standard eclipse (Devillard \cite{eclipse}) and IRAF procedures for data reduction. The available twilight flats were used to create a bad pixel map. We removed bad pixels, ghosts and cosmic rays in all frames before combining them. Except for one object (J1143-8036c) we used dome flats for flat fielding. An illumination correction to these flat fields was done to remove a small residual gradient in the sky. Residual bias subtraction was only necessary for the objects with the highest quality data (J0738+0507b and J1143-8036d). The spectral tilt was removed using star traces. The OH lines were used to correct for the curvature of spectral lines. If the detected emission line was close to one of the edges of the detector, we recalculated this correction optimising it for the area around the emission line to minimise OH line residuals where they are most relevant. 

Flux calibration was done using bright B stars, observed directly after the object. A (wavelength dependent) slit loss correction was applied to the 0\farcs6 slit spectra. The OH lines were used for wavelength calibration, $<0.5$ \AA ~residuals remained after a third order fit.

 One or more emission lines are immediately visible in the two dimensional spectra in 5 out of 9 cases. In two cases, we also detect continuum emission, in one case, we detect continuum emission without an emission line (J1056-0337). One detection turns out to be a Seyfert 1 (J0931-0449). The reduced two dimensional spectra of the detected emission lines are shown in Fig. \ref{spec2D} (except the Seyfert).

\begin{figure*}
   \centering
\includegraphics[width=5cm]{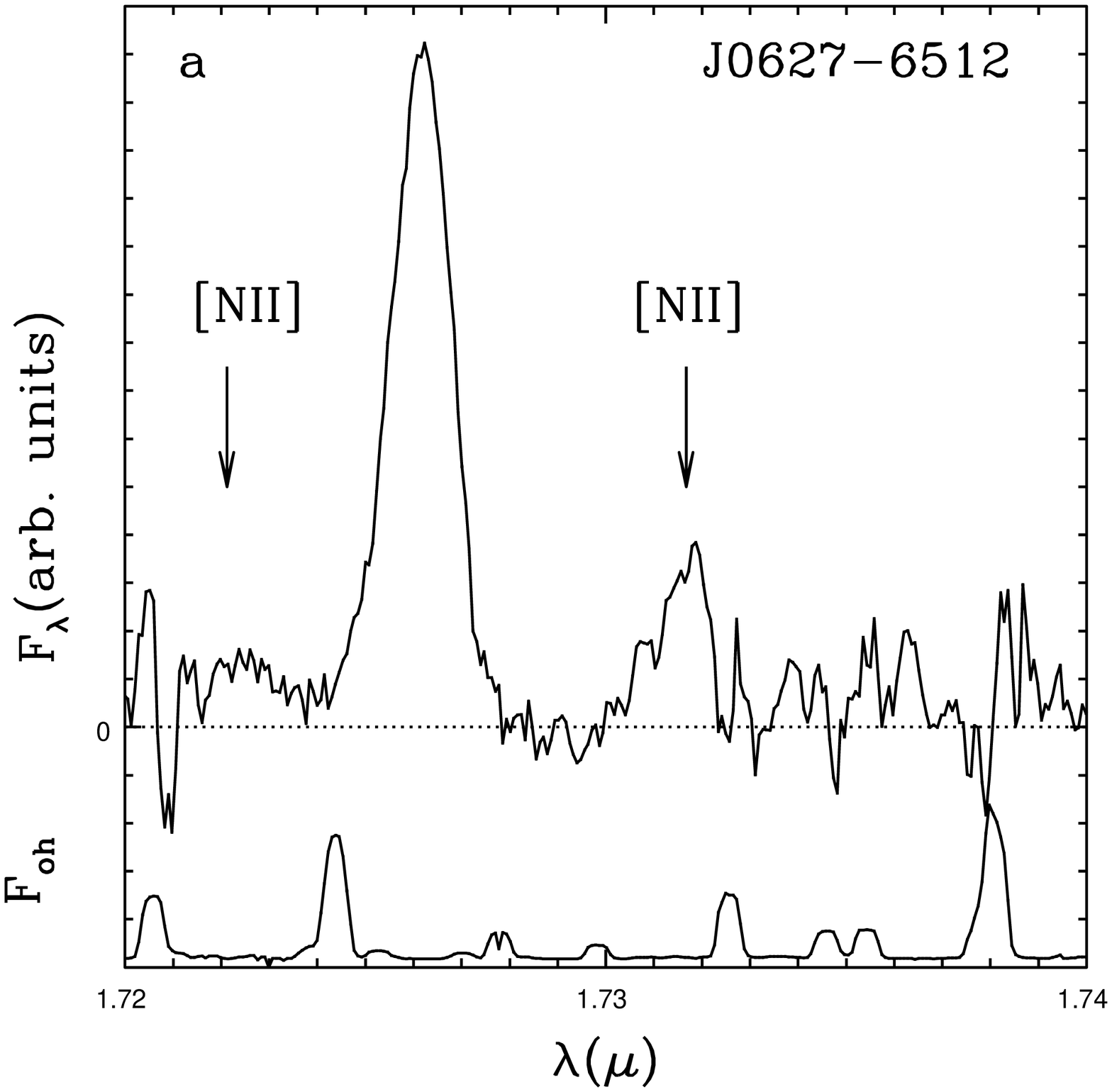}
\includegraphics[width=5cm]{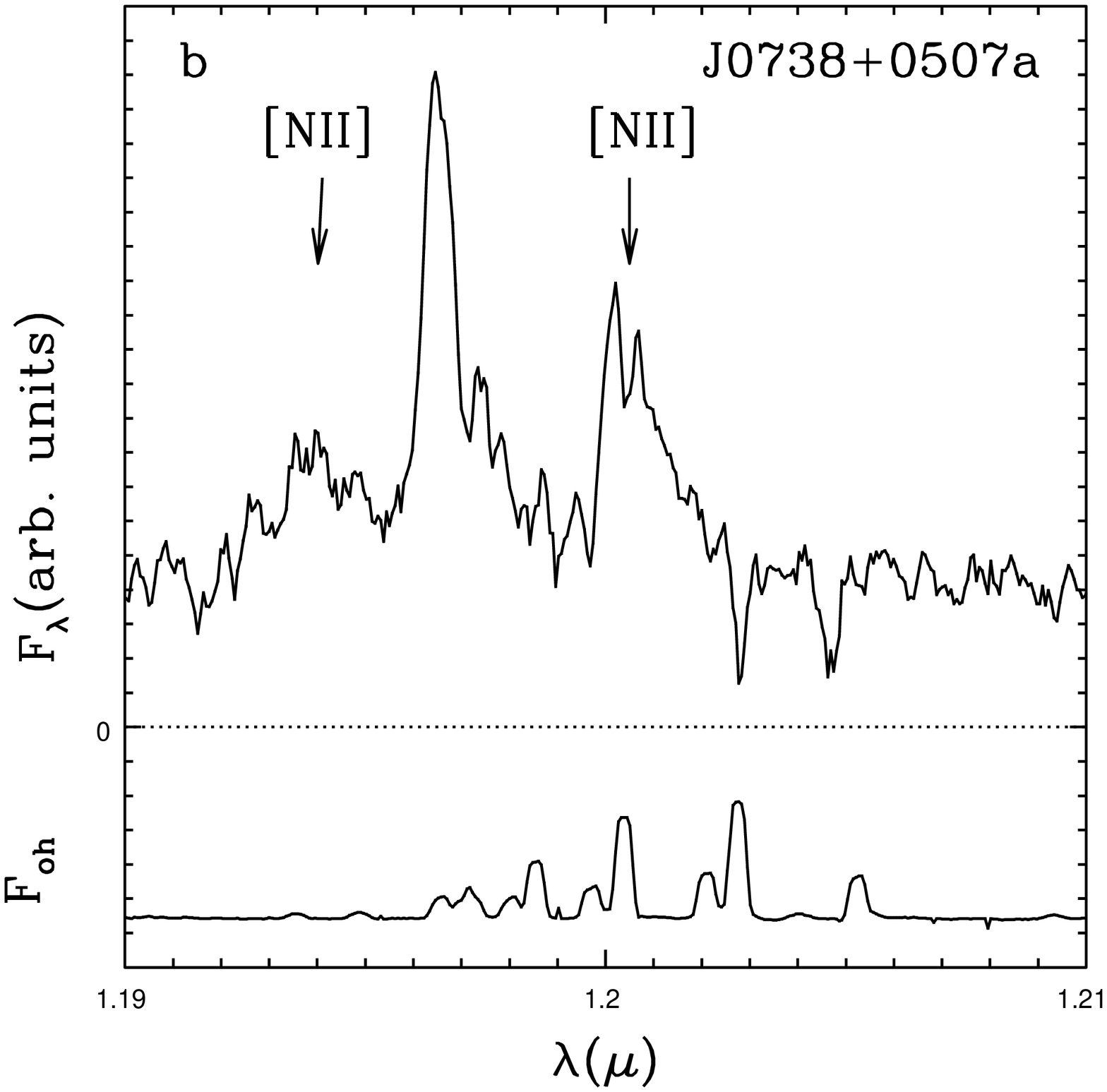}
\includegraphics[width=5cm]{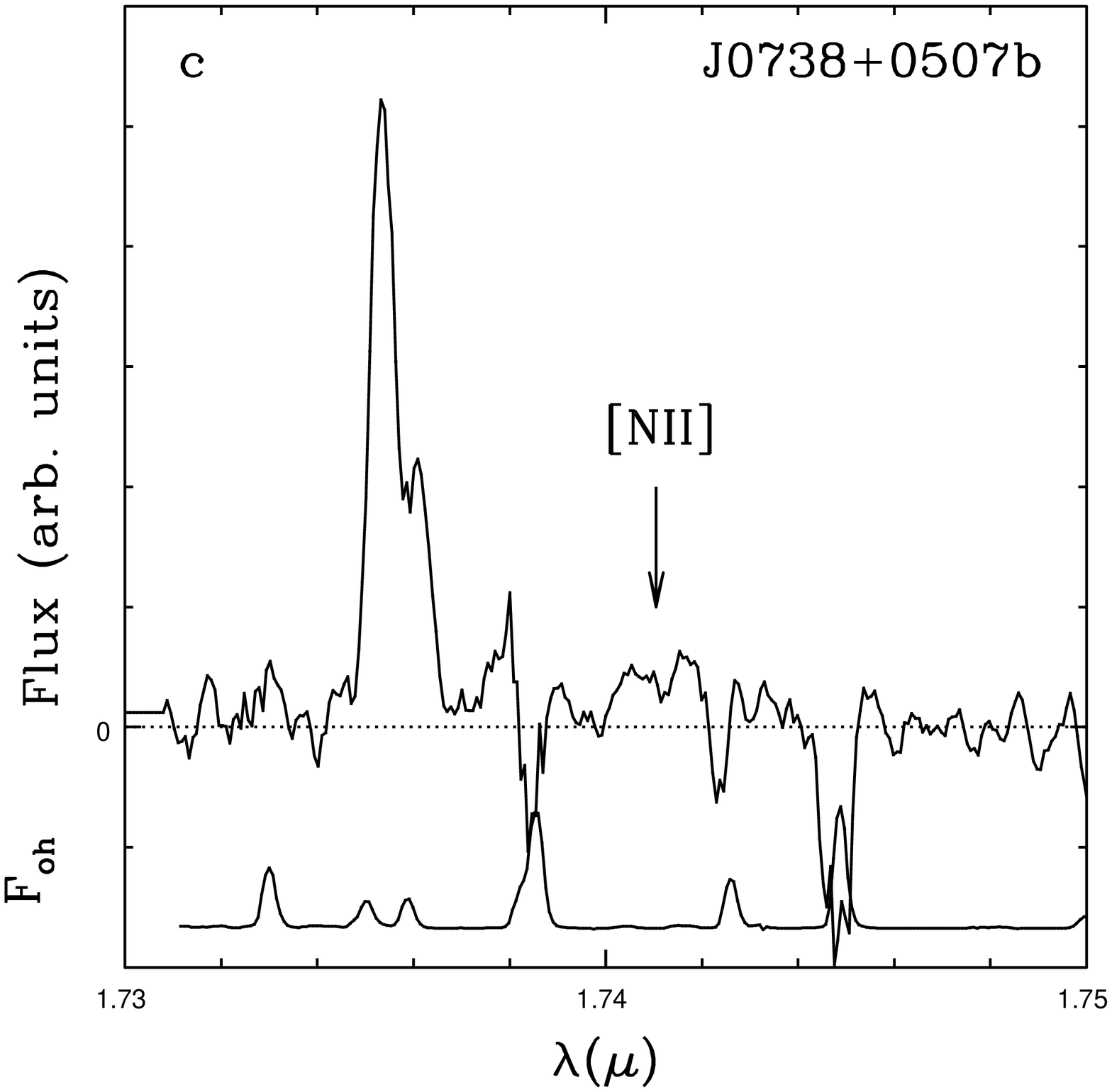}
\includegraphics[width=5cm]{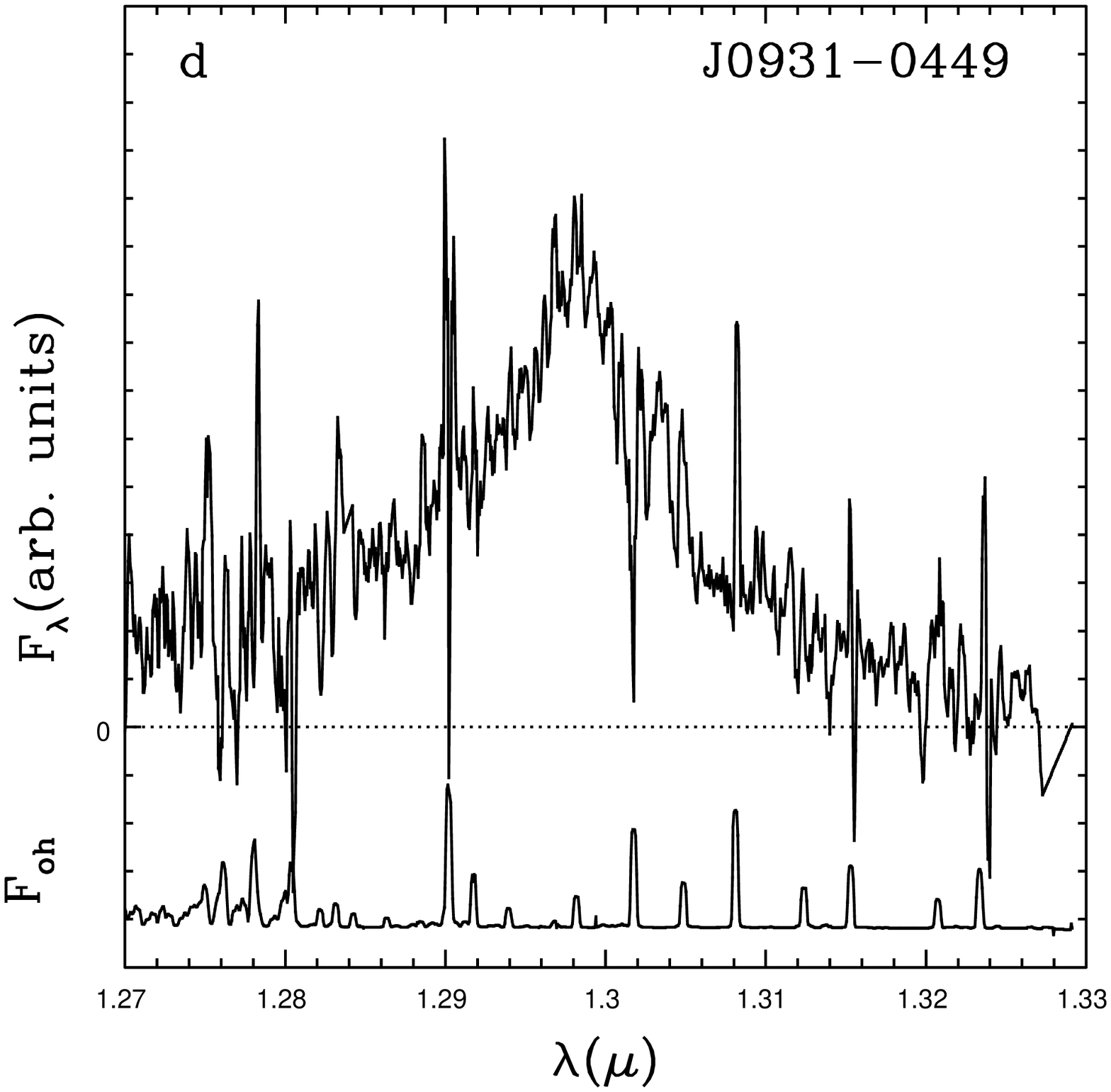}
\includegraphics[width=5cm]{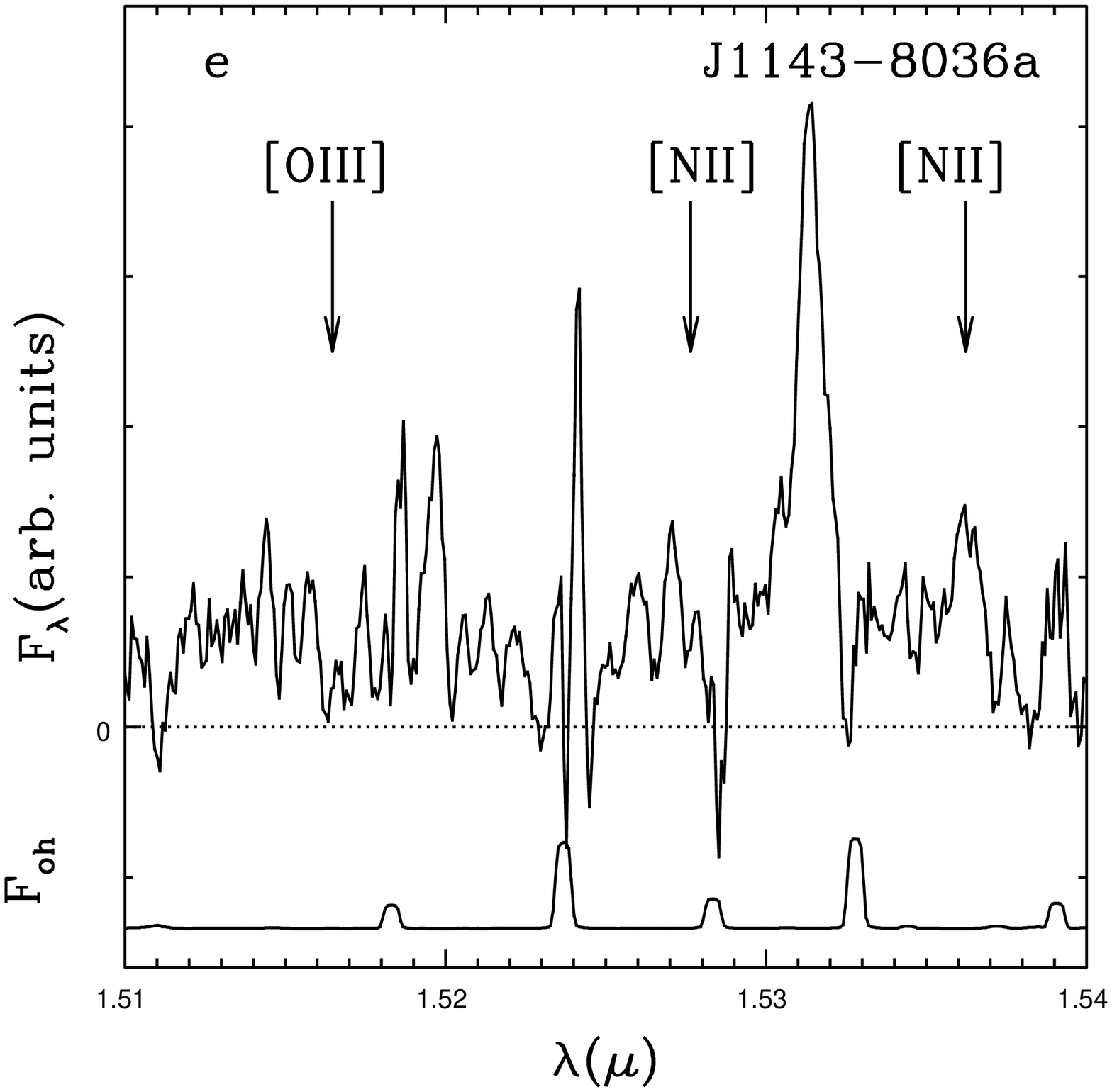}

      \caption{1D spectra for all sources. Assuming the brightest emission line detected is H$\alpha$, the expected position of the [\ion{N}{ii}]6584/6548 lines are marked. If [\ion{N}{ii}]6584/6548 is not or marginally detected, we also marked the expected position of [\ion{O}{iii}]4959 assuming the brightest emission line is [\ion{O}{iii}]5007. Note that [\ion{O}{iii}]4959 and [\ion{N}{ii}]6548 fall outside the wavelength range observed for object J0738+0507b.
              }
         \label{spec1D}
   \end{figure*}

We extracted one dimensional spectra by cutting out a strip from the two dimensional spectrum containing all flux of the detected emission line, or, if there was no (clear) detection in the 2D spectrum, a strip was cut out at the expected position of the emission line (using the known distance between the object and the reference star). Extracting the spectrum by tracing the spectrum was not an option, because we detect weak continuum emission in only three sources. The spectra were smoothed with a Gaussian with FWHM approximately equal to the FWHM of the OH lines in the raw frames and are shown in Fig.~\ref{spec1D}. In the one dimensional spectra, a second or third emission line is immediately evident in two cases. The brightest emission line of every detected object was clearly visible in the two dimensional spectrum. 

Line fluxes and widths were measured by fitting a Gaussian to the detected emission lines using IRAF's `splot'. Errors were estimated by repeated fitting with different parameters. If there was severe OH line contamination, we interpolated over the OH line to correct for flux losses. Error bars are naturally larger in this case. Linewidths were corrected for instrumental broadening and converted to $W_{20}$, the width at 20\% of maximum flux. Line fluxes and widths are given in Tables \ref{emlin} and \ref{sfrmass} respectively. Values corrected for OH line contamination are marked by the subscript 'OH'.

   \begin{table*}
      \caption[]{Emission lines detected}
         \label{emlin}
     $$ 
         \begin{array}{llllll}
            \hline
            \noalign{\smallskip}
            \mbox{source ID}      &  \mathrm{\lambda(\mu m)} & \mbox{line ID} & z &  F^{\mathrm{a}} & F_\mathrm{OH}^{\mathrm{a}\mathrm{b}}  \\
            \noalign{\smallskip}
            \hline
            \noalign{\smallskip}
            \mbox{J0627--6512}  & 1.72616  & \mathrm{H\alpha} & 1.630  &  4.0  \pm 0.2  &    \\
	    \mbox{J0627--6512}  & 1.73167  & [\ion{N}{ii}]6584 & 1.630  &  0.59 \pm 0.05 &      \\
	    \mbox{J0738+0507a}  & 1.1966   & \mathrm{H\alpha} & 0.824  &  2.9  \pm 0.3  & \sim 4.1 - 4.7     \\
	    \mbox{J0738+0507a}  & 1.2005   & [\ion{N}{ii}]6584 & 0.824  &  2.0  \pm 0.2  & \sim 2.3 - 3.7     \\
	    \mbox{J0738+0507a}  & 1.19411  & [\ion{N}{ii}]6548 & 0.824  &  1.6  \pm 0.2  &       \\
	    \mbox{J0738+0507b}  & 1.7354   & \mathrm{H\alpha} \mbox{ or } [\ion{O}{iii}]5007 & 1.644 \mbox{ or } 2.466 & 0.74 \pm 0.02 & \sim 0.9 - 1.0     \\
	    \mbox{J0931--0449}  & 1.2973   & \mathrm{H\alpha} + [\ion{N}{ii}]6548/6584 & 0.977  &  21 \pm 5 &      \\
	    \mbox{J1143--8036a} & 1.53137  & \mathrm{H\alpha} & 1.333  &  0.73 \pm 0.06 &       \\
	    \mbox{J1143--8036a} & 1.53623  & [\ion{N}{ii}]6584 & 1.333  &  0.16 \pm 0.03 &      \\
            \noalign{\smallskip}
            \hline
         \end{array}
     $$ 
\begin{list}{}{}
\item[$^{\mathrm{a}}$] Fluxes in units of $10^{-16}\textrm{erg s}^{-1} \textrm{cm}^{-2}$.  
\item[$^{\mathrm{b}}$] F$_\mathrm{OH}$ is an estimate of the emission line flux F had it not been contaminated by one (or more) OH sky lines.

\end{list}
   \end{table*}

   \begin{table*}
      \caption[]{SFRs and masses.}
         \label{sfrmass}
     $$ 
         \begin{array}{lllllll}
            \hline
            \noalign{\smallskip}
            \mbox{Source ID}    &  \mathrm{SFR} & \mbox{SFR}_\mathrm{OH} & \mathrm{W}_{20} & \mathrm{W}_{20 \mathrm{ ~OH}} & \mbox{M} & \mathrm{R}^{\mathrm{a}}  \\
	                        & \mathrm{M}_{\sun} \textrm{yr}^{-1}   & \mathrm{M}_{\sun} \textrm{yr}^{-1} & \mathrm{km s}^{-1} & \mathrm{km s}^{-1}  & 10^{10} \mathrm{M}_{\sun} & \mathrm{kpc} \\
            \noalign{\smallskip}
            \hline
            \noalign{\smallskip}
            \mbox{J0627--6512}               & 57 \pm 3     &              & 344 \pm 11   &             & 4.8 &  7\\
	    \mbox{J0738+0507a}               & 20 \pm 2     & \sim 28 - 32 & 398 \pm 38   & \sim 479-677& 5.1 &  6 \\
	    \mbox{J0738+0507b}^{\mathrm{b}}  & 10.9 \pm 0.3 & \sim 13 - 15 & 166 \pm 3    & \sim 216-235& 1.3 &  8\\
	    \mbox{J0931--0449}^{\mathrm{c}}  & 167 \pm 40   &              & 5300\pm 1800 &             &     &   \\
	    \mbox{J1143--8036a}              & 8.2 \pm 0.7  &              & 274 \pm 18   &             & 1.7 &  4\\
            \noalign{\smallskip}
            \hline
         \end{array}
     $$ 
\begin{list}{}{}
\item[$^{\mathrm{a}}$] R is half the diameter, measured as the total extent along the slit in the spectrum. 
\item[$^{\mathrm{b}}$] SFR and mass were calculated assuming the emission line observed is H$\alpha$. 
\item[$^{\mathrm{c}}$] \mbox{J0931--0449} is a Seyfert 1 galaxy, the SFR is not meaningful. 
\end{list}
   \end{table*}

We calculated rest frame B or R magnitudes (depending on redshift) from the observed F110W (J hereafter) and F160W (H hereafter) magnitudes. An H band magnitude was available for all targets, J band for a subset only. When J band photometry was unavailable, we used the average J-H color (equal to the median color) of the entire NICMOS grism sample to estimate the J band magnitude. We calculated the rest frame magnitudes by interpolating between the J and H fluxes, depending on redshift this gives us a rest frame B or R absolute magnitude. Where we could not interpolate to obtain a rest frame B or R magnitude, we made a rough estimate of this magnitude by using the closest flux point available. The errors in the absolute magnitudes were calculated as follows: while interpolating between the J and H magnitudes, interpolating meaning that the redshifted effective wavelength of the B or R band lies in between the effective wavelengths of the J and H band, we set the error in the measured magnitudes to 0.1, and in the J magnitudes calculated from the average J-H color to 0.4 (=scatter in J-H color).  We then interpolated the fluxes and uncertainties to get the rest frame magnitudes and errors.  When the effective wavelength of the redshifted B or R band was not between the effective wavelengths of the J and H band but inside the wavelength range of the J or H band, we set the error to 0.5 mag. If it was outside the wavelength range of the J and H band, we set the error to 2.0 mag. 
These numbers are a bit arbitrary, but are intended to reflect the increased uncertainties in the magnitudes estimates.  The apparent infrared magnitudes and corresponding absolute rest frame magnitudes are given in Table \ref{mag}.  When calculating the offsets from the TFR, we use only those points where the redshifted B or R filter at least overlapped with the observed J or H band.

  \begin{table}
      \caption[]{Apparent magnitudes and absolute magnitudes of targets}
         \label{mag}
     $$ 
         \begin{array}{lcccc}
            \hline
            \noalign{\smallskip}
            \mbox{Source ID}     &  \mbox{J}  & \mbox{H}  & \mbox{M}_{\mbox{B}}  & \mbox{M}_{\mbox{R}} \\
            \noalign{\smallskip}
            \hline
            \noalign{\smallskip}
            \mbox{J0627--6512}  &  22.2    &  20.4  &  	-22.3 \pm 0.1    &   (-23.1 \pm 0.5) \\
	    \mbox{J0738+0507a}  &  <18.8>  &  17.9  &   (-22.5 \pm 2~~)~ &   -23.7  \pm 0.1  \\
	    \mbox{J0738+0507b}  &  <22.9>  &  22.0  &   -20.7 \pm 0.1    &   (-21.5 \pm 0.5) \\
\mbox{J0738+0507b}([\ion{O}{iii}])  &  <22.9>  &  22.0  & -21.7 \pm 0.1  &   (-22.9 \pm 2~~)~ \\
	    \mbox{J0931--0449}  &  19.7    &  19.0  &   (-22.2 \pm 2~~)~ &   -23.2  \pm 0.1 \\
 	    \mbox{J1143--8036a} &  <22.3>  &  21.4  &   (-20.6 \pm 0.5)  &   -21.4 \pm 0.1	\\
 
            \noalign{\smallskip}
            \hline
         \end{array}
     $$ 
\begin{list}{}{}
\item[$^{\mathrm{a}}$] J and H band magnitudes from McCarthy et al. (\cite{mccarthy}). The J magnitudes calculated from the H band magnitude and average J-H color are marked by $<~>$. The extrapolated absolute magnitudes and their errors (see text) are in parentheses.
\end{list}
   \end{table}

\section{Results}

We detect one or more emission lines in 5 out of 9 spectra. One dimensional spectra (of the relevant wavelength ranges of the original ISAAC spectra) are shown in Figs. \ref{spec1D}a-e. In these figures, the expected positions of the [\ion{N}{ii}]6584/6548 lines (assuming the brightest emission line detected is H$\alpha$) are marked. If the observed emission line is not H$\alpha$, the next most likely candidate is [\ion{O}{iii}]4959/5007. Also marked is the expected position of [\ion{O}{iii}]4959 (assuming the bright line detected is [\ion{O}{iii}]5007) if the detection of the [\ion{N}{ii}] doublet is uncertain. To avoid confusion between emission lines and OH line residuals, the sky spectra are also shown. Other possible identifications of the emission lines can be ruled out or are far less likely: [\ion{O}{ii}]3727 would put the sources at redshifts larger than 3 (H band detection) and would be resolved in a doublet which is not observed. The equivalent width of H$\beta$ is in general too low to be detected in the McCarthy et al. (1999) survey. We find 4 H$\alpha$ emitting galaxies and 1 (likely) [\ion{O}{iii}]5007/4959 emitting galaxy. In Sect. 5, we discuss all galaxies individually.

In Table \ref{mccarthy}, we list the wavelengths and fluxes from McCarthy et al. (\cite{mccarthy}). We note that there is a systematic offset between the wavelength as found by McCarty et al. (\cite{mccarthy}) and ours, although all our wavelengths lie within $3\sigma$ error bars of the NICMOS wavelengths. We checked some of the OH lines in the ISAAC spectra and they were correct within a few \AA. We also note that the emission line fluxes are not always in agreement. This is probably due to a combination of the low resolution of the NICMOS grism and slit losses with ISAAC. 

  \begin{table*}
      \caption[]{Comparison with the line fluxes and wavelengths of McCarthy et al.}
         \label{mccarthy}
     $$ 
         \begin{array}{lllllll}
            \hline
            \noalign{\smallskip}
            \mbox{Source ID} & \lambda_{McC}(\mu m) & \lambda(\mu m)  & \Delta\lambda(\AA) & F_{McC}^{\mathrm{a}} & F^{\mathrm{a}} & F_{OH}^{\mathrm{a}}\\
            \noalign{\smallskip}
            \hline
            \noalign{\smallskip}
            \mbox{J0627--6512}  & 1.742  & 1.72616  & 158  & 1.8 \pm 0.5  &  4.0  \pm 0.2  &  \\
	    \mbox{J0738+0507a}  & 1.210  & 1.1966   & 134  & 16  \pm 1.5  &  2.9  \pm 0.3  & \sim 4.1 - 4.7 \\
	    
	    \mbox{J0738+0507b}  & 1.77   & 1.7354   & 346  & 0.9 \pm 0.3  & 0.74 \pm 0.02 & \sim 0.9 - 1.0 \\
	    \mbox{J0931--0449}  & 1.299  & 1.2973   & 17   & 24  \pm 1.7  &  21 \pm 5 &  \\
	    \mbox{J1143--8036a} & 1.538  & 1.53137  & 66   & 1.2 \pm 0.4  &  0.73 \pm 0.06 &  \\
            \noalign{\smallskip}
            \hline
         \end{array}
     $$ 
\begin{list}{}{}
\item[$^{\mathrm{a}}$] All fluxes are in units $10^{-16} \textrm{erg s}^{-1} \textrm{cm}^{-2}$.
\end{list}
   \end{table*}

In Table \ref{sfrmass} we list starformation rates (SFRs) and dynamical masses for all detected objects. SFRs were calculated using 
\begin{equation} \textrm{SFR(M}_\odot \textrm{yr}^{-1}) = \textrm{L}_{H\alpha}(\textrm{erg s}^{-1}) / 1.26 \times 10^{41} \end{equation}
(Kennicutt et al. (\cite{kennicutt}) for a Salpeter Initial Mass Function (IMF)).

Dynamical masses were calculated using \begin{equation}M(R) =  \frac{RV^2}{G}\end{equation} where the velocity $V=W_{20}/2$ and the diameter $R=D/2 $. $D$ is the diameter of the galaxy measured as the total extent in the spectrum. We also measured the diameters in the images with gave consistent results. The masses and radii are also listed in Table \ref{sfrmass}. Note that these masses are lower limits as no correction for inclination or OH lines has been applied.

The [\ion{N}{ii}]/H$\alpha$ ratio can be used to get an estimate of the metallicity of galaxies. We used the calibration of Denicol\'o et al. (\cite{denicolo}) and the results are reported in Table \ref{metallicity}.

   \begin{table}
      \caption[]{Metallicity}
         \label{metallicity}
     $$ 
         \begin{array}{llll}
            \hline
            \noalign{\smallskip}
            \mbox{Source ID}    &  [\ion{N}{ii}]/\mbox{H}\alpha & \log([\ion{N}{ii]}/\mbox{H}\alpha ) &  12 + \log(\mbox{O/H})     \\
            \noalign{\smallskip}
            \hline
            \noalign{\smallskip}
	    
	    \mbox{J0627--6512}   &    0.125 \pm 0.020 &  -0.90 \pm 0.16 &  8.46 \pm	0.16 	\\	
	    \mbox{J0738+0507a}^{\mathrm{a}}   &	0.69  \pm 0.14	&  -0.16 \pm 0.20 &  9.00 \pm	0.16 	\\	
	                  &	\la 1   \pm 0.1 	&   \la 0    \pm 0.1  &  \la 9.12 \pm   0.09\\
	    \mbox{J0738+0507b}   &    \la 0.005^{\mathrm{b}} \pm 0.005	&  \la -2.3  \pm   1  &   \la 7.4 \pm    0.8 	 \\	
	    \mbox{J1143--8036a}  &	0.22  \pm 0.06	&  -0.66 \pm 0.27 &  8.64 \pm   0.22	\\	

            \noalign{\smallskip}
            \hline
         \end{array}
     $$ 
\begin{list}{}{}
\item[$^{\mathrm{a}}$] The OH line corrected values are on the second row. Note this galaxy is probably a narrow line AGN and the line ratio cannot be interpreted as a metallicity effect.
\item[$^{\mathrm{b}}$] This upper limit is based on the bright part of the emission line, measured in 2D image (to get best constraint)
\end{list}
   \end{table}

\section{Notes on individual objects}
We will now discuss all galaxies individually, paying attention to the identification of the emission line(s), H$\alpha$/[\ion{N}{ii}] ratios, linewidths and kinematics. Where we do not detect any emission line, we will attempt to give an explanation. 

\paragraph{J0627-6512} A single bright emission line is visible between two bright OH lines (see Figs. \ref{spec2D}a and \ref{spec1D}a). Although [\ion{N}{ii}]6584 emission is not visible by eye in the two dimensional spectrum, it is quite obvious in the one dimensional spectrum. Hence, we confirm the redshift to be 1.630. The [\ion{N}{ii}]6584/H$\alpha$ ratio is about 0.13, confirming that we are looking at a star forming galaxy (Brinchmann et al. 2003, Gallego et. 1997). We do not detect continuum emission in the spectrum.

In the NICMOS image, this object looks like an asymmetric edge-on galaxy. Indeed, we do detect some extended emission in the spectrum on the same side of the galaxy as in the image. Remarkably, the emission line is not tilted, and there is no sign of ordered rotation. The elongated appearance of the galaxy could be intrinsic, and not due to an edge-on orientation. We might also miss a tilt in the emission line due to the observation conditions, see Sect. \ref{velocity} for a discussion of this possibility. The optical seeing during these observation varied between 0\farcs65 and 1\farcs24 (median seeing 0\farcs75).  With current data, we cannot distinguish between the two possibilities. We would need excellent seeing, better $S/N$ (the $S/N$ of the extended emission is poor) data with a smaller slit to determine the nature of J0627-6512.

\paragraph{J0738+0507a} This is a very bright source, both in emission line and in continuum flux. We detect H$\alpha$, [\ion{N}{ii}]6548 and [\ion{N}{ii}]6584 and continuum emission at redshift 0.823. Because of its brightness and its compact morphology, it has been suggested that J0738+0507a hosts an Active Galactic Nucleus (AGN) (McCarthy et al. \cite{mccarthy},  Hicks et al. \cite{hicks}). However, our detection of narrow emission lines rules out the possibility of a Seyfert 1. Star forming galaxies and AGNs can be distinguished from their emission line ratios due to their different excitation properties. The most suited line ratio diagram to separate star forming galaxies from AGNs is the line ratio diagram with $\log([\ion{N}{ii}6584]/\textrm{H}\alpha)$ on one axis and $\log([\ion{O}{iii}]/\textrm{H}\beta)$ on the other axis (e.g. Brinchmann et al. \cite{brinchmann}). We do not have measurements of all these lines, but the ratio $\log([\ion{N}{ii}]/\textrm{H}\alpha)$ can identify some (but not all) AGNs. According to Brinchmann et al. (\cite{brinchmann}), all galaxies with $\log([\ion{N}{ii}]/\textrm{H}\alpha) > -0.2$ are AGNs. The measured $\log([\ion{N}{ii}]/\textrm{H}\alpha)$ ratio for J0738+0507a is quite uncertain, because both lines are contaminated by OH line emission. The measured value (see Table \ref{metallicity}) is $-0.16 \pm 0.20$. The true value is most probably larger (less negative), because the [\ion{N}{ii}] line is more contaminated then the H$\alpha$ line. We suggest that J0738+0507a is likely a narrow line AGN.

\begin{figure}
   \centering
\includegraphics[width=8.5cm]{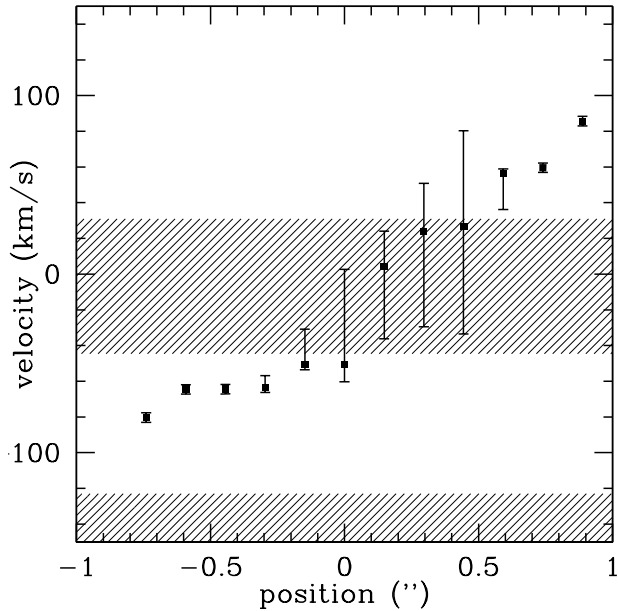}

      \caption{Rotation curve of J0738+0507b. The shaded areas correspond to OH lines. }
         \label{rotcurve}
   \end{figure}

\paragraph{J0738+0507b} This galaxy was observed under excellent atmospheric conditions with the 0\farcs6 slit and a total integration time of just over 4 hours. McCarthy et al. (1999) refer to it as a ``putative emission line''. We detect a beautifully tilted emission line, extending $\sim 1\farcs8$ (1\arcsec corresponds to $\sim$ 8 kpc at the source redshift, see below) and rising continuously without flattening off, see Fig. \ref{spec2D}c and Fig. \ref{rotcurve}. The rotation curve velocity $2V$ is at least $\sim110 \textrm{ km s}^{-1}$ (total visible extent of the emission line without correction for inclination or the OH line cut off (see below)).

The wavelengths of the OH sky lines turn out to be very unfortunate: one bright OH line cuts off the detection on the short wavelength side, another falls on the middle of the emission line, making it hard to put strong limits on emission line flux, extent and velocity. Strictly speaking we can only measure lower limits.

What is clear however, is that we do not observe a double horned profile. What we observe is a bright part and a much fainter part on the long wavelength side. It is possible that a similar fainter outer part is also present on the short wavelength side, but this is impossible to detect due the the presence of the bright OH line at that side of the emission line. Comparing the flux as a function of position in the spectrum to the flux in the image, and assuming that the equivalent width of the emission line does not vary with position, we checked that there is some continuum flux in the image at the undetectable position in the spectrum. We could therefore be looking at a centrally star bursting system, with lower levels of star formation in the outer parts. 

As can be seen in Fig. \ref{spec1D}c, there is no sign of [\ion{N}{ii}]6584 and there is no bright OH line close to the expected position of [\ion{N}{ii}]6584. We can rule out an H$\alpha$/[\ion{N}{ii}]6584 ratio smaller than 190 at the 3$\sigma$ level (for the brightest part of the emission line, assuming constant H$\alpha$/[\ion{N}{ii}]6584 for the whole galaxy).
The highest H$\alpha$/[\ion{N}{ii}]6584 ratios observed for local starburst galaxies are $\sim 20$ for Blue Compact Dwarfs with some outliers with $\sim 100$ (Gallego et al.1997, Brinchmann et al. 2003), therefore identification of the emission line as H$\alpha$ seems unlikely. 

Unfortunately, the detected emission line is close to the edge of the detector. Both [\ion{N}{ii}]6548 (assuming H$\alpha$) and [\ion{O}{iii}]4959 (assuming [\ion{O}{iii}]5007) fall off the detector, prohibiting confirmation of [\ion{O}{iii}]5007/4959. Although McCarthy et al. (1999) could in principle have resolved the [\ion{O}{iii}] doublet (the separation between the lines at this redshift would be about twice their resolving limit), the small line flux immediately explains why they could not in this case. We conclude that [\ion{O}{iii}]5007 is the most likely candidate, putting the redshift at 2.466 (instead of 1.644) and making this object one of the highest redshift objects with ordered rotation. We note that high redshift [\ion{O}{iii}] emitting galaxies have been misidentified as H$\alpha$ before: Moorwood et al. (\cite{moorwood}) find that most of their presumed H$\alpha$ emitters are [\ion{O}{iii}]5007/4959 emitters.  

\paragraph{J0931-0449}   
This galaxy is identified as an AGN: it has a very broad emission line (FWHM $\sim 2500$ km s$^{-1}$), confirmed to be H$\alpha$ (and [\ion{N}{ii}]6548 and [\ion{N}{ii}]6584) by Hicks et al. (\cite{hicks}) who detected [\ion{O}{ii}]3727. 

\paragraph{J1056-0337}
The emission line of this galaxy was not detected, although the emission line flux reported by McCarthy et al. (\cite{mccarthy}), $4.6 \times 10^{-16} \textrm{erg s}^{-1} \textrm{cm}^{-2} $, is by far not the faintest in the sample and we do detect continuum emission. Possible explanations are sky line contamination (parts of the spectra are very crowded with OH lines), extended emission, or a spurious source in the NICMOS sample (J1056-0337 is the lowest redshift object from the NICMOS sample, the emission line is near the edge of the wavelength range covered by the NICMOS grism, the signal-to-noise of the detection is only 3). This object has also been observed as part of the FIRES survey of MS1054-03 (F\"orster-Schreiber et al. \cite{forster}), and has photometric redshift of 0.4 (Franx, private communication) whereas the redshift according to McCarthy et al. is 0.72. 

\paragraph{J1143-8036a}
This source has the lowest S/N detection in our sample. We see a tentative detection of [\ion{N}{ii}]6584, although we cannot rule out other possibilities. Continuum emission is not detected.

\paragraph{J1143-8036b}
J1143-8036a and J1143-8036b were detected by McCarthy et al. (\cite{mccarthy}) as a pair with nearly identical redshifts. J1143-8036b was the fainter one of the pair, and here we barely detect J1143-8036a (the distance to the reference star of the detected emission line ruled out the other interpretation: non-detection of J1143-8036a, detection of J1143-8036b). As J1143-8036b has lower emission line flux than J1143-8036a, and J1143-8036a was barely detected, J1143-8036b is probably below the detection limit.

\paragraph{J1143-8036c}
This galaxy was not detected. The emission line may fall below the detection limit or the line may fall on top of a bright OH line. 

\paragraph{J1143-8036d}
This galaxy was not detected in the spectrum despite excellent observing conditions and long integration times. We cannot explain this.
   
\paragraph{J1120+2323a} This galaxy was scheduled in service mode under excellent conditions, but could not be identified on the acquisition image, is spite of repeated attempts. \\

We confirm line emission in 5 out of 9 galaxies, including one Seyfert 1, one possible Seyfert 2 and three starburst galaxies. In one case, the emission line is most likely [\ion{O}{iii}]5007 instead of H$\alpha$. The results in this paper are therefore based on a very small number of sources (three), one of them having an uncertain redshift.

\section{The Tully-Fisher relation at $z > 1$}
We will now turn to the discussion of high redshift TFRs, using the NICMOS galaxies as a case study. We will first use this data set to present the $z\sim1.5$ starburst TFR without any corrections whatsoever. Then, we discuss extinction and inclination corrections for high redshift galaxies. We will then explain our choices for the velocity parameter, the luminosity parameter and the local comparison sample, and how the results would change if other choices had been made. Finally selection effects in velocity, magnitude, sample specifics and star formation are discussed. The whole discussion is strongly focused on the pitfalls of high redshift TFR analysis in order to assess what can be attributed to an evolving TFR and what to peculiarities of this and other samples.

\subsection{The $z \sim 1.5$ starburst TFR}
For the case study, we study the rest frame B and R band starburst TFR using $W_{20}$ on the velocity axis. As a local reference sample, we take the Verheijen (\cite{verheijen}) B and R band TFR with again $W_{20}$ on the velocity axis. This sample has been corrected for inclination and extinction (extinction correction recipe from Tully et al. \cite{tully}). The Verheijen sample consists of spiral galaxies form the Ursa Major cluster and is among the tightest TFRs in the literature. The velocities are HI measurements.

\begin{figure*}
   \centering
\includegraphics[width=8.5cm]{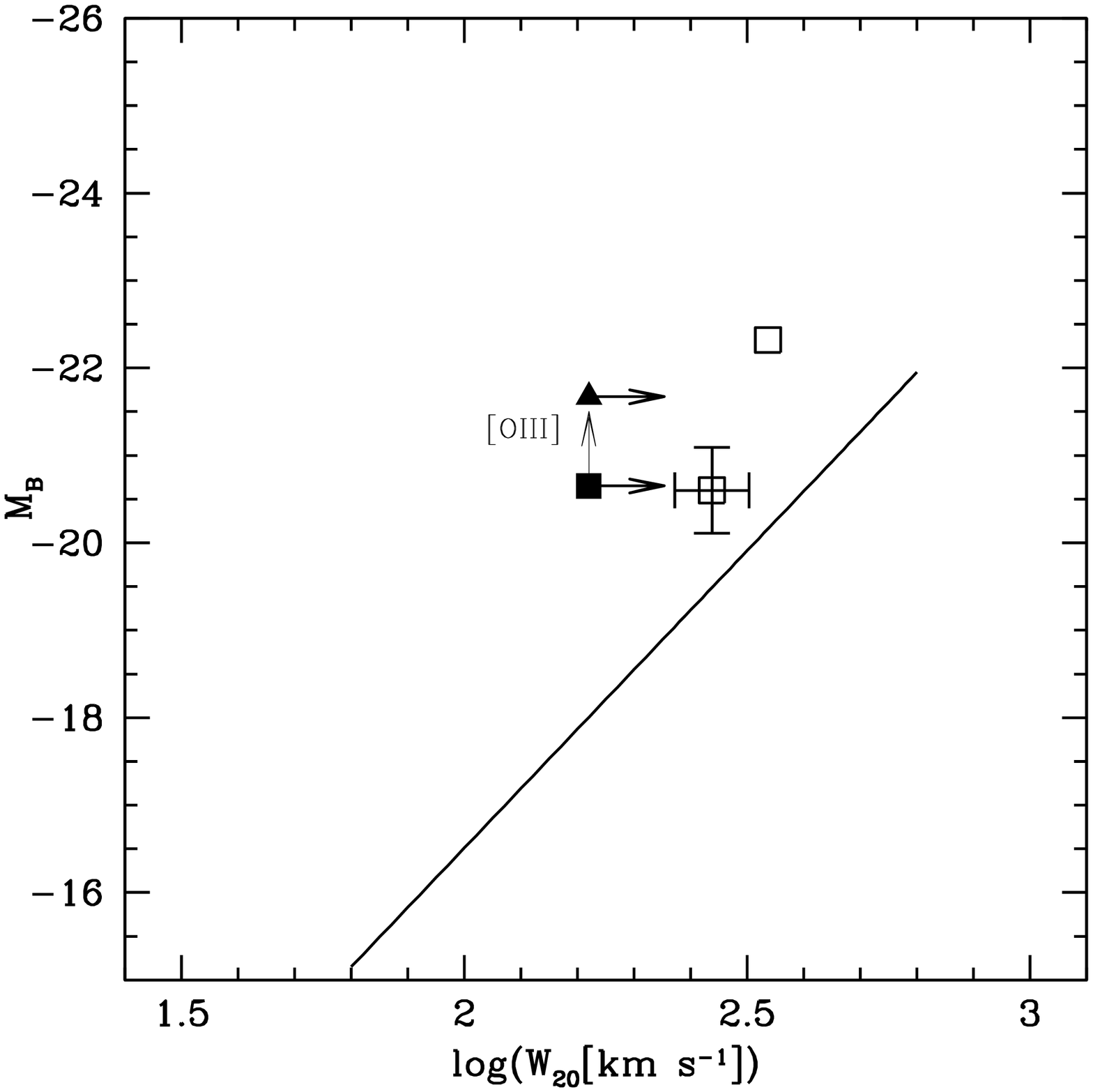}
\includegraphics[width=8.5cm]{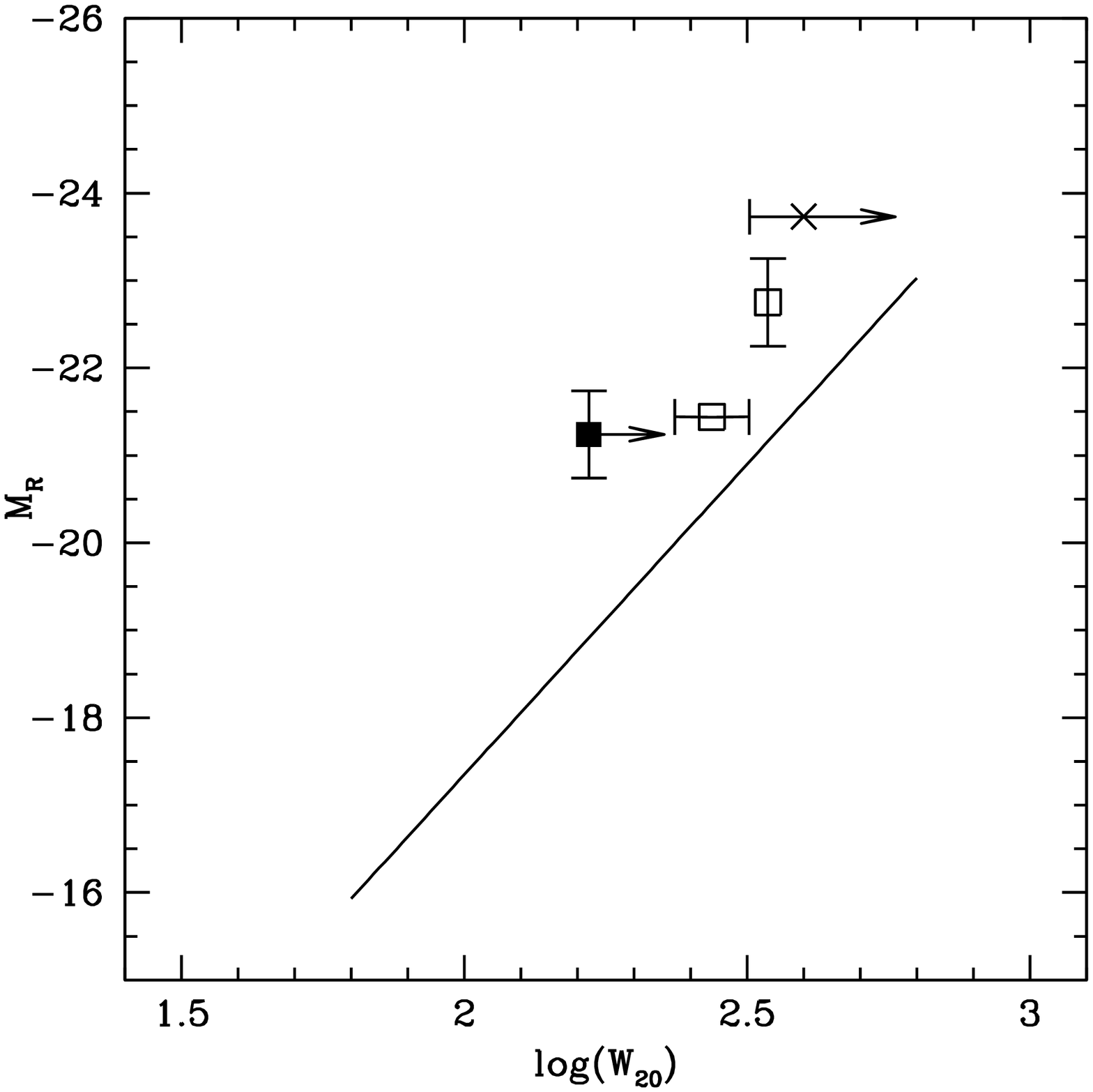}

      \caption{The local B band (left figure) and R band (right figure) TFR of Verheijen (\cite{verheijen}) with over plotted the $z \sim 1.5$ NICMOS galaxies.  The two solid symbols both correspond to J0738+0507b for different redshift identifications. They are connected by an arrow to make this clear.    The triangle is J0738+0507b assuming the detection emission line is [\ion{O}{iii}] instead of H$\alpha$ (this point is not in the R band TF plot because its rest frame R magnitude is highly uncertain, see Table \ref{mag}). J0738+0507a is presumably a Seyfert 2 galaxy and is marked by a cross. The arrows indicate an estimate of underestimation of the linewidth due to influence of nearby OH lines.  As discussed in Sect. \ref{velocity} the linewidths are subject to a large number of uncertainties. For example, if a correction for beam smearing could be made, all linewidths would shift even further to the right.  If the error bars were comparable or smaller than the point sizes, they were left out for reasons of clarity. 
              }
         \label{tfplotbr}
   \end{figure*}

In Figs. \ref{tfplotbr}a and b we plot the local B and R band TFR of Verheijen (\cite{verheijen}) with our redshift $\sim1.5$ objects, which were not corrected for inclination or extinction. All points lie significantly above the local TFR. Without any corrections (and excluding the possible narrow line AGN J0738+0507a), we estimate that the $z=1.5$ TFR lies $\sim$ 2.0 magnitudes above the local TFR in B and $\sim$ 1.8 magnitudes in R at $\log (W_{20}[\textrm{km s}^{-1}]) \sim 2.5$. For comparison, using ISAAC in a similar way, Barden et al. (\cite{barden}) found an offset of around 1 magnitude at $z\sim1$.

\subsection{Effects of extinction and inclination}
Contrary to the local reference sample used (and most local TFRs), our data has not been corrected for extinction and inclination. We did not attempt to correct our data points for extinction or inclination because both corrections are very uncertain for our data. Not correcting for extinction means that we measure a lower limit to the offset from the TFR.

Correcting velocities (line widths) for inclination would decrease the measured luminosity offset. To bring the sample on the local TFR, the required shift in $\log(W_{20})$ is 0.29 in B and 0.25 in R. Assuming random inclinations, the average shift in $\log (\textrm{W} _{20})$ would be 0.3. Therefore, the lower limit to the luminosity offset is $\sim 0$.

\begin{figure}

\includegraphics[width=8.5cm]{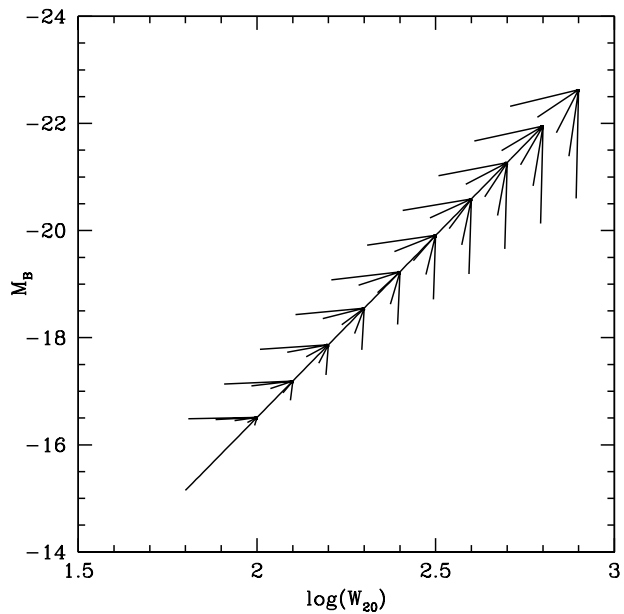}

      \caption{The combined effect of extinction and inclination. Plotted are the local TFR of Verheijen (\cite{verheijen}) with the vectors showing where a galaxy lying on the TFR would lie if it had not been corrected for inclination and extinction (according to Tully et al. \cite{tully}) for inclinations 80\degr, 70\degr, 60\degr, 50\degr and 40\degr from bottom to top. 
              }
         \label{extincl}
   \end{figure}

However, the effects of extinction and inclination are not independent: more inclined galaxies are more heavily extincted. Depending on inclination and velocity width (if the extinction correction is velocity dependent like the Tully et al. (\cite{tully}) correction), the slope of the shift in position between the location of the uncorrected and corrected point in the TF plot, is steeper or shallower than the slope of the (local) TFR, and the measured luminosity offset increases or decreases respectively. For large inclinations (edge-on galaxies), the net effect is a lower limit on the luminosity offset. In Fig. \ref{extincl}, we plot the combined effect of extinction and inclination correction using the (local) correction recipe of Tully et al. (\cite{tully}). Plotted are the local TFR from Verheijen (\cite{verheijen}) and vectors showing where points on the TFR would lie if they had not been corrected for extinction and inclination for inclinations of 80, 70, 60, 50 and 40\degr. The figure shows that for galaxies at the relevant velocity range, we measure a lower limit to the luminosity offset if they are more inclined than  $\sim60\degr$. Increasing the amount of extinction (and/or a shallower local TFR, the Verheijen TFR is among the steepest known) lowers this turnover inclination. One magnitude extinction more than locally lowers the turnover inclination to $\sim40\degr$. For a sample of galaxies with random inclinations, the net effect on the measured luminosity offset is then $\sim 0$. Although we cannot measure the inclinations accurately, we can say that J0627-6512, J0738+0507b and J1143-8036a are significantly inclined, given their elongated morphology even at limited spatial resolution. Therefore, we are confident that the measured offsets are lower limits to the luminosity offset, although the precise amount remains uncertain due to the small number of sources.

Of course this result relies on the assumption that local extinction corrections apply for the galaxies in our sample, which is uncertain. Under this assumption however, the measured offset from the local TFR is robust against the combined effect of inclination and extinction. 

Accurate inclinations for high redshift galaxies can be obtained with high resolution imaging. Without any extinction correction, the measured offsets from the local TFR are then lower limits by definition. If high resolution imaging is not available, the above approach minimises uncertainties. For all high redshift TFR samples, one has to assume local extinction corrections apply for high redshift galaxies unless an independent estimate of the extinction is available. The Balmer decrement or SED fitting could constrain the extinction, although the latter method suffers from degeneracy between age and dust.

\subsection{Velocity: W$_{20}$}\label{velocity}

A number of different parameters may be plotted on the velocity axis of the TF-plot. Examples for rotation curves are the maximum rotation curve velocity $V_{\rm{max}}$ and the velocity of the flat outer part of the rotation curve $V_{\rm{flat}}$. When using linewidths (double horned profiles) one usually takes the width at some percentage (50\% or 20\%) of the maximum flux or average flux of both horns. A number of possible corrections may be applied to those linewidths, for example for instrumental broadening and mass motions. All TFR velocity parameters correlate well with eachother. The tightest  TFRs are found using $V_{\rm{flat}}$, which traces best the total mass of the galaxy, on the velocity axis (Verheijen \cite{verheijen}). 

We use the width at 20\% of maximum flux W$_{20}$ as an approximation for twice the flat rotation curve velocity $V_{\rm{flat}}$. We correct the measured linewidth for instrumental broadening. We choose W$_{20}$ as our line width definition because in local Tully-Fisher samples, the difference between $2V_{\rm{flat}}$ and W$_{20}$ is smallest (compared with other line width definitions).

Although there are some prescriptions to convert the HII $W_{20}$ linewidth to velocity or HI linewidth (see Rix et al. \cite{rix}, Pisano et al. \cite{pisano}), we did not apply any of those as they give contradicting corrections factors. Pisano et al. (\cite{pisano}) show for a sample of nearby blue compact galaxies that the correction factor from $W_{20}(\textrm{HII})$ to $W_{20}(\textrm{HI})$ is large ($\sim20\% $ or more) for galaxies with small linewidths ($ W_{20}(\textrm{HII}) \la 140 \textrm{ km/s}$) and small for galaxies with large linewidts (less then 10\% for galaxies with linewidths comparable to ours). On the other hand, they also find that for galaxies with equivalent width comparable to ours, the HII linewith is up to a factor 2 smaller than the HI linewidth. 
Rix et al. (\cite{rix}) cite different studies investigating the relation between optical linewidths and flat circular velocity. They conclude that $W_{20}(\textrm{HII})$ should be corrected upward by 14\% regardless of $W_{20}(\textrm{HII})$, equivalent to reducing the luminosity offset by 0.2 mag in B band for the slope of the Verheijen (\cite{verheijen}) TFR. 
We conclude that the uncertainty in the correction factor is significant although the correction in $\log(W_{20})$ (and hence luminosity offset) is probably small. Therefore, we decide not to correct the $W_{20}$ linewidth (except for instrumental broadening).

A crucial assumption in our analysis is that the velocity widths and the rotation curve velocity (in the case of J0738+0507b) trace the mass of the galaxy in a similar way as in the nearby universe. That is, that linewidth traces the ordered rotation of the outer parts of the galaxy where the rotation velocity is constant. This assumption is highly non-trivial to prove (or disprove). Both observational effects and fundamental questions underly this assumption.

Inflows, outflows, mergers and other dynamical disturbances may bias the line width low {\it or} high. Slit effects may prohibit the identification of peculiar kinematics. More uncertainties arise due to slit spectroscopy: the kinematic and photometric major axes need not coincide, and the velocity profile is smeared out because the slit and the galaxy are of comparable size. Poor spatial resolution can bias the RC velocity low up to a factor two (Erb et al. \cite{erb04}). Poor spectral resolution causes smearing in the spectral direction. Poor signal-to-noise, $S/N$, observations cause a significant uncertainty in the linewidths and RC velocity observed (Moorwood et al. \cite{moorwood}).  The combined effect on the measured line widths is uncertain and cannot be quantified with our current data, leaving significant uncertainties in the measured offset from the TFR.

The distribution of the emission line flux over the galaxy also influences the linewidths and RC velocity measurements. Barton \& Van Zee (\cite{barton}) showed that a central (or more general, a local) peak in star formation may bias the linewidth measured low by a factor 2. 

A more fundamental question is: do the optical emission lines extend out to the flat part of the RC like they do in the local universe? Although tilted emission lines have been observed out to redshift 3.2 (Moorwood et al. \cite{moorwood}; Pettini et al. \cite{pettini}), no flat RCs have been observed above redshift $\sim 1 $ (Moorwood et al. \cite{moorwood}; Pettini et al. \cite{pettini}; Lemoine-Busserolle et al. \cite{lemoine}; van Dokkum \& Stanford \cite{vandokkum}; Erb et al. {\cite{erb03}, \cite{erb04}). A large sample of $z > 1$, high $S/N$, high spectral resolution and excellent seeing observations may show these disks do exist at these redshifts, but it may also be the case that star forming disks have undergone significant evolution since $z \sim 1$ and do not extend to the flat part of the RC. P\'erez (\cite{perez}) recently reported evidence for stellar disk truncation in the redshift range $0.6 < z < 1.0$.

Kannappan et al. (\cite{kannappanprep}) investigated the importance of various effects on the offset from the local TFR of two high redshift TFR studies ($z \sim 0.34$ and $\sim 0.52$). When all samples are converted to the same cosmology and reference sample, the largest correction factor is that of rotation curve truncation (0.71 magnitude using their cosmology and reference sample). Their samples used RCs only, the situation for linewidths might be different, nonetheless this should be a major warning that our velocity measurements cause a large uncertainty in our analysis.

\subsection{Rest frame magnitude}
TFRs can be measured in all optical and near infrared bands. The scatter in the TFR decreases with longer wavelength because of decreased sensitivity to dust and star formation (e.g. Verheijen \cite{verheijen}). Our choice for rest frame B and R band was forced by the available photometry. Spitzer will open the window to rest frame K band photometry. 

\subsection{Local comparison sample}\label{localcomp}
One can only measure a luminosity offset with respect to a reference sample. The galaxies in the reference sample may be and probably will be different from the high redshift galaxies in many respects. They may have different SFRs and star formation histories (SFH), higher ages, different metallicities and dust properties, different emission line fluxes and EWs, more or less or other kinematic disturbances, lower gas mass fractions, and more. All these factors contribute directly or indirectly to the position of the galaxies in the TF plot and the luminosity offset measured. Careful definition of the reference sample is crucial for the interpretation for the results. In our case, we choose to take the Verheijen (\cite{verheijen}) sample as a reference sample. This sample contains only 'very ordinary' cluster spirals. Strictly speaking, we have measured the offset of young star forming galaxies from the 'most ordinary local spirals' TFR. We will show now that our conclusions do not change if we choose an other reference sample.

A local starburst sample might be an obvious choice for a reference sample. Several local starburst samples for TFR applications are available and they follow the normal local TFR, with outliers and increased scatter mainly at the low mass end (Mendes de Oliviera et al. \cite{mendes}; Barton et al. \cite{bartongeller}; Van Driel, Van den Broek \& Baan \cite{vandriel}; Brungardt \cite{brungardt}; Davoust \& Contini \cite{davoust}). The effects of star formation on the TFR become apparent at $\log(W_{20})\la2.4$. When studying high mass galaxies (like we do), taking a starburst TFR does not make a difference for the measured luminosity offsets. We discuss the effects of star formation on the TFR extensively in subsection \ref{TFstarburst}.

We might also choose a sample containing all local spirals, following Kannappan et al. (\cite{kannappanprep}) who used a morphology blind sample of bright emission line galaxies brighter than $M_B = -18 $ and inclined more than 40 degrees. This kind of TFR generally has a shallower slope than other TFRs, because of the larger effect (see subsection \ref{TFstarburst}) of a starburst on lower mass galaxies. The approach also has it pitfalls, as the high redshift sample does not contain 'all' high redshift galaxies, and depending on the high redshift sample, the amount of luminosity evolution may appear different. Using this TFR to measure the offset from the local TFR at $\log(W)=\log(2V)=2.5$ reduces the measured offset by 0.2 mag (Kannappan et al. relation) plus 0.15 mag (to account for the different value used for H$_0$) = 0.35 mag.

Our sample is selected on EW(H$\alpha$) and F(H$\alpha$), a local sample selected on these properties could also serve a reference sample. Unfortunately, there  is no such sample available.

We conclude that the differences between local TFRs for different samples of galaxies are important for the low mass end of the TFR. The differences for high mass end of the TFR are negligible compared to the other uncertainties in the analysis.

\subsection{Selection effects}\label{seleffects}

Selection effects play an important role in high redshift TFR studies. We will now discuss selection in velocity, selection in magnitude, and selection in equivalent width. The latter is specific for our sample. 

In Fig. \ref{selvel}, we plot once and twice the for formal slit limits for ISAAC velocity resolution (1\arcsec ~slit, MR mode R $\sim3000$) for H$\alpha$ at redshift 1.5. These limits vary very little between redshift 1 and 2.5 for H$\alpha$ measurements with ISAAC. These limits are upper limits to the actual resolution because seeing is the limiting factor. It is immediately evident why measuring TF slopes is so difficult at high $z$: the range in velocity is small compared with local samples, where velocity resolution limits only play a role in samples of dwarfs. Therefore, we did not attempt to measure a 'relation' but only an offset from the local TFR at $W_{20} \sim  320 \textrm{ km s}^{-1}$ corresponding to  $\log (W_{20}[\textrm{km s}^{-1}]) \sim 2.5$.

 \begin{figure}

\includegraphics[width=8.5cm]{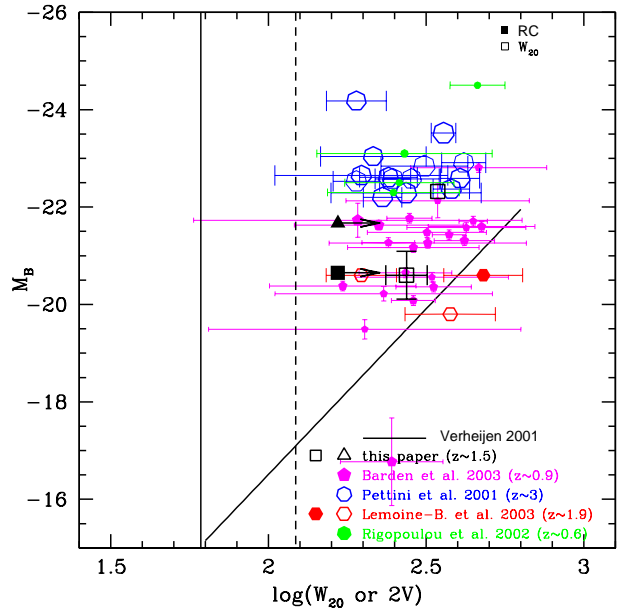}

      \caption{Selection in velocity for the high redshift TFR. Plotted are the local TFR of Verheijen (\cite{verheijen}), the data points presented here (using the same symbols as in Fig. \ref{tfplotbr}) and data from various other authors whose data were also (partially) obtained with ISAAC. Filled symbols are RC velocities, open symbols linewidths and point size increases with redshift. The vertical lines are once (thick line) and twice (dashed line) the ISAAC velocity resolution (1\arcsec slit MR mode) for H$\alpha$ at redshift 1.5. 
              }
         \label{selvel}
   \end{figure}

 It becomes increasingly difficult to measure TFR parameters for galaxies of a certain luminosity with increasing redshift, pushing one again to larger mass (and hence brighter) galaxies. There is no clear absolute magnitude cut in the NICMOS grism sample, it is biased against low equivalent width emission lines (see below). Moreover, the redshift range is quite large and the depth of the observations varies from field to field. To give an impression of a typical luminosity limit, the magnitude limit for the lowest luminosity object in our sample (in the total NICMOS grism sample there are a few slightly fainter objects) at redshift 1.5 would be $-20.4$ in B. At the velocities observed, we cannot distinguish between a zero point offset or increased scatter in the TFR in magnitude limited samples.

The main selection effect in the NICMOS grism sample is a bias against emission lines with low equivalent width. A selection effect on EW cannot be translated to a simple straight line in the TF plot, like the selection effect on velocity and magnitude. However, the selection effect on EW could be very important. Kannappan et al. (\cite{kannappan}) showed that in the local universe EW(H$\alpha$) correlates with offset from the TFR, with large EW(H$\alpha$) galaxies being up to 2 mag brighter in R than expected from the local TFR although with considerable scatter. Kannappan \& Barton (\cite{kannappanbarton}) showed a similar result for the B band residuals. However, there are three reasons why these results should not be simply copied to our galaxies.

First, the galaxies in those samples have rest frame EW(H$\alpha$) $\la 60$ \AA, and most have EW$\sim 20$ \AA. The rest frame EW(H$\alpha$) of our sample of galaxies ranges from 161 to 310 \AA ~(detections only, and excluding the Seyfert 1) (numbers from McCarthy et al. \cite{mccarthy}), i.e. an order of magnitude larger. A local sample of galaxies with comparable EW(H$\alpha$) and L(H$\alpha$) line strengths is unavailable due to the rarity of these galaxies in the nearby universe (James et al. \cite{james}; Gallego et al. \cite{gallego}). Based on the EW(H$\alpha$) alone, we might expect the NICMOS galaxies to be brighter than the not EW selected $z\sim1.5$ TFR although the amount of brightening remains uncertain.

Second, Kannappan et al. (\cite{kannappan}) and Kannappan \& Barton (\cite{kannappanbarton}) also showed that B-R color correlates with residuals from the B and R band TFR, with the residuals being about $\sim$0.5 mag at the estimated B-R color for the NICMOS galaxies. This amount of brightening is significantly less than expected from the EW(H$\alpha$), indicating again we are looking at an 'incomparable sample'.

Third, the systematic offset from the TFR correlating with EW(H$\alpha$) and B-R color as found Kannappan \& Barton (\cite{kannappanbarton}) mainly comes from galaxies with $\log$(W[km s$^{-1}$]) $\la$ 2.4 (their definition of line width W is different from ours, but the differences between different definitions of W are minor). The galaxies in Kannappan et al. (\cite{kannappan}) cannot easily be traced through their diagrams. EW(H$\alpha$) may be important for low mass galaxies, but irrelevant for high mass galaxies, and hence the NICMOS galaxies. 

We therefore conclude that although the selection of high EW galaxies most likely biases our result, the effect cannot be quantified by a comparison to local samples. We therefore take a different approach in the next section.

\subsection{The effect of a starburst on luminosity in the TF plot}\label{TFstarburst}

The study of the high redshift TFR is limited to galaxies with a sufficient SFR to detect and resolve emission lines like H$\alpha$, [\ion{O}{iii}]5007/4959 and [\ion{O}{ii}]3727. These galaxies may not be representative (from a TFR point of view) for galaxies with smaller SFR. Star formation may affect the position of a galaxy in the TF plot in both luminosity and velocity. The effects on velocity were discussed in Sect. \ref{velocity}. A starburst increases the luminosity of a galaxy, and one might worry that the use of only starburst galaxies will mimic luminosity evolution in the TFR. We will now show that the increase in $\log(\textrm{L})$ (the parameter in the TF plot) is negligible for massive galaxies assuming they have a significant older population, which - as we will show below - might not be the case for some of the NICMOS galaxies.

We already discussed a hint that star formation could be important in Sect. \ref{seleffects}, namely the results of Kannappan et al. (\cite{kannappan}, \cite{kannappanprep}) and Kannappan \& Barton (\cite{kannappanbarton}). These papers showed that offsets from the U, B and R band TFR correlate with the star formation indicators B-R color and global EW(H$\alpha$), bluer color and larger EW (hence more actively star forming) galaxies being overluminous for their RC velocity. 
We also noted that different parameters gave conflicting results for our sample, and outliers mainly occur at the low mass end of the TFR. 
On the other hand, there was also a hint that star formation may not be important for TFR studies. As we saw in Sect. \ref{localcomp}, local starburst TFR follow the local normal spiral TFR, with some outliers at the low mass end. We will now try to explain these observations using a simple model.

We plot in Fig. \ref{starburst} the effect on luminosity of local galaxies for different starbursts (different SFRs, different duration of the starburst) calculated using models from Bruzual \& Charlot (\cite{bruzual}, (BC)) (Padova 1994 models, Cabrier IMF, solar metallicity). We added only the luminosity of the starburst, and neglected the effect of the additional mass on the RC velocity. As can be seen in this figure, the luminosity of low mass galaxies is dominated by even modest amounts of additional star formation. For galaxies with $\log(W_{20}) \ga 2.4$ only strong starbursts have  a significant effect on $\log(L)$. This is the $W_{20}$ value where the BC model (see Fig. \ref{starburst}) for a short ($10^7 $ yr) $10\textrm{ M}_{\sun}\textrm{ yr}^{-1} $ starburst lies 1 mag above the (local) TFR. This figure immediately explains why outliers are often low mass galaxies and why the zero point of the TFR at the high mass end is so stable.

 \begin{figure}

\includegraphics[width=8.5cm]{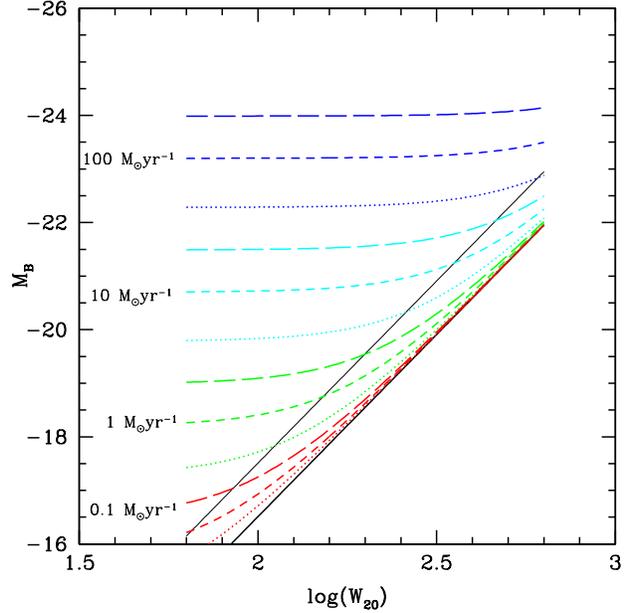}

      \caption{The luminosity of starbursts of different SFRs (indicated in the plot) and different durations ($10^7$, $10^8$ and $10^9$ yr for the dotted, dashed and long dashed lines respectively) added to the local Verheijen (\cite{verheijen}) TFR (thick line). The thin line indicates one mag brightening.
              }
         \label{starburst}
   \end{figure}

High redshift galaxies have a younger and less massive population compared with local galaxies. The net effect on luminosity using BC models with an exponentially decreasing SFR is an increase in $L_B$ for high redshift galaxies. Additional starformation will therefore have a smaller effect on the luminosities of high redshift galaxies compared to local galaxies.

We conclude that provided the older stellar population is sufficiently luminous and the measured velocity represents the mass of the galaxy, a starburst will not change the position of a galaxy in the TF plot and starburst galaxies are (from a TFR point of view) representative for the total galaxy population. 

We will now attempt to check if the NICMOS galaxies fulfill these criteria and are therefore TFR-representative for the entire $z\sim 1.5$ spiral galaxy population. The linewidths indicate that the NICMOS galaxies are sufficiently massive for TFR studies: their linewidths (without inclination correction) are $\log(W)\ga 2.4$ . The non-detection of rotation curves can be explained by poor seeing. Improving the velocity measurements remains the biggest challenge for all high redshift Tully-Fisher studies.

For the case of the NICMOS galaxies, the luminosity of the older population is hard to constrain. The SFRs indicate we cannot assume the starburst luminosity to be negligible.  The large EWs indicate the importance of a young stellar population. We used the starburst99 models (Leitherer et al. \cite{starburst99}) to estimate the age of the galaxies. Assuming constant star formation and a Salpeter IMF with mass range $1 - 100 \textrm{M}\odot $, the EW indicate that the age of the galaxies is $\la 1$ Gyr. This results strongly depends on the initial mass function and star formation history. A younger starburst could also produce the large EW, while contributing less to the total luminosity. Without additional data, we cannot calculate a reliable estimate of the starburst luminosity, and we cannot answer the question whether this sample is TFR-representative for the entire $z\sim1.5$ spiral galaxy population. Note that this conclusion is specific for this sample and not for other samples at similar redshift. The EWs of this sample are extreme, and we have no photometric data that could provide constraints on the SFH.

\section{Summary and conclusion}

We studied the challenges in measuring $z > 1$ TFR using emission line galaxies using a sample of H$\alpha$ emitting galaxies. We conclude:

- We confirm only 5 out of 9 emission lines found by McCarthy et al. (\cite{mccarthy}). Of the 5 remaining sources, one is a broad line AGN, one probably a narrow line AGN. Our conclusions are therefore based one a very small number of sources. 

- Without reliable inclinations and extinctions, an ensemble averaged simultaneous inclination and extinction correction can be made assuming the relation between extinction and inclination follows that of local galaxies, which is in general a small correction because galaxies move approximately along the TFR. 

- The most difficult challenge for high redshift TFR studies, are the velocity measurements. Besides the observational challenges and limitations, there is also the fundamental question if what we observe is comparable to what we observe in the local galaxies (for example the question if the emission lines extent to the flat part of the RC). For the current data set, this leaves considerably uncertainties in the measured line widths, which cannot be quantified.

- The high mass end of the local TFR is not sensitive to the sample used, therefore the choice of the local reference sample is not important for the high mass end of the high redshift TFR. 

- Extending the study of the high redshift TFR to low mass galaxies will be very difficult due to the selection effects in velocity and magnitude: it would require very high spectral resolution observations of very faint galaxies.

- Star formation increases the luminosity of galaxies, but the effect on $\log(L)$ is negligible for high mass galaxies with significant older population. Therefore star forming galaxies can be TFR-representative for all galaxies, and whether they are or not can be checked.

For the NICMOS galaxies, we measured a $\sim 2$ mag offset of the $z \sim 1.5$ rest frame B starburst TFR with respect to the local TFR. This offset is robust against the effects of inclination and extinction. The linewidths indicate sufficiently massive galaxies for TFR studies, the results is therefore also robust for the choice of local comparison sample. However, we cannot prove or disprove that the all linewidths are due to rotation. Moreover,  we cannot answer the question if the star burst luminosity dominates the total luminosity or not. This is due to the extreme nature of the galaxies and the sparseness of data for this sample.

\begin{acknowledgements}
We thank the ESO-Paranal staff for executing the service mode observations.

\end{acknowledgements}

\end{document}